\documentclass[aps,pra,twocolumn,longbibliography]{revtex4-1}
\usepackage{amsmath}
\usepackage{amssymb}
\usepackage{MnSymbol}
\usepackage{lineno}
\usepackage{ulem}

\newcommand\beq{\begin{equation}}
\newcommand\eeq{\end{equation}}
\newcommand\bea{\begin{eqnarray}}
\newcommand\eea{\end{eqnarray}}
\newcommand\nn{\nonumber}

\newcommand{\ket}[1]{| #1 \rangle }

\newcommand\aop{\hat a}
\newcommand\phit{\tilde{\phi}}
\newcommand\phiop{\hat{\phi}}
\newcommand\xv{\mathbf{x}}
\newcommand\kv{\mathbf{k}}
\newcommand\ki{{\kappa i}}
\newcommand\al{\alpha}
\newcommand\om{\omega}
\newcommand\Hop{\hat H}

\begin{document}
\title{No, classical gravity does not entangle quantized matter fields}
\author{Lajos Di\'osi}
\email{diosi.lajos@wigner.hu}
\affiliation{Wigner Research Center for Physics, H-1525 Budapest 114 , P.O.Box 49, Hungary}
\affiliation{E\"otv\"os Lor\'and University, H-1117 Budapest, P\'azm\'any P\'eter stny. 1/A, Hungary}
\date{\today}

\begin{abstract}
In their recent work, Nature, {\bf 646}, 813 (2025), Aziz and Howl claim that classical (unquantized)
gravity can generate entanglement of quantized matter if matter is treated within quantum field theory
which is, no doubt, our ultimate theory to use. We show that the perturbative result 
of Aziz and Howl in interaction picture is inconsistent with our exact and simple non-perturbative 
derivation in Heisenberg picture, that fundamentally precludes the claimed entanglement.
\end{abstract}
\maketitle


In their recent work \cite{aziz2025classical},
Nature, {\bf 646}, 813 (2025), Aziz and Howl claim that classical (unquantized)
gravity can generate entanglement between quantized matter if matter is treated within quantum field theory,
which is, no doubt, our ultimate theory to use. This surprising result would imply that experimental tests of
gravity's quantumness are 'harder than previously thought' \cite{weller2025there}. The authors argue
that classical gravity causes the exchange of virtual particles between separate parts of the
second-quantized matter, leading to entanglement between the parts. We show that, under their own assumptions,
the model of Aziz and Howl is equivalent to free quantum field dynamics on a fixed background metric, 
which is solvable exactly in the Heisenberg picture. The perturbative result of Aziz and Howl is inconsistent
with our exact and simple non-perturbative derivation that fundamentally precludes the claimed entanglement.

Ref. \cite{aziz2025classical} considers the 
initial state of a complex boson field of mass $m>0$
in the product (unentangled) form:
\bea
\ket{\Psi}=\frac12&&\Bigl(\ket{N}_{1L}\ket{0}_{1R}+\ket{0}_{1L}\ket{N}_{1R}\Bigr)\nonumber\\
                    \otimes&&\Bigl(\ket{N}_{2L}\ket{0}_{2R}+\ket{0}_{2L}\ket{N}_{2R}\Bigr).
\eea
The four N-boson states, labelled by $\ki$, with $\kappa\in\{L,R\},i\in\{1,2\}$, 
are four spatially separated Fock-states over the four local vacuum states:
\beq
\ket{N}_\ki=\frac{1}{\sqrt{N!}}(\aop_\ki^\dagger)^N\ket{0}_\ki.
\eeq
The four annihilation operators are of the standard forms:
\beq
\aop_\ki=\int\frac{d^3\kv}{(2\pi)^3}\phi_\ki(\kv)\aop_\kv,
\eeq
where $\phi_\ki(\kv)$ are the Fourier transforms of four non-relativistic spatially separated
single-boson normalized wave functions $\phit_\ki(\xv)$, respectively. Using linearized
semiclassical Einstein equations, ref. \cite{aziz2025classical} claims that coupling to classical gravity turns
the product state (1) into an entangled state $\ket{\Psi(t)}$. The claim is based on 4$^{th}$ order perturbative
terms. This claim contradicts the consensus that semiclassical gravity cannot generate entanglement.

We are going to examine eq. (1).  Uncorrelated (product) forms of distant local vacua or their excited Fock states
are approximate forms in field theory, restricted to the local domains in question, and valid under suitable
conditions only. Fortunately,  they have their exact, unconditionally valid forms.
The exact form of the  product state $\ket{N}_{1L}\ket{0}_{1R}$, for instance,
is $(N!)^{-1/2}(\aop_{1L}^\dagger)^N\ket{0}$, where $\ket{0}$ is the standard vacuum.
Accordingly, the exact form of eq. (1) must be
\bea
\ket{\Psi}=\frac{1}{2N!}&&\Bigl((\aop_{1L}^\dagger)^N+(\aop_{1R}^\dagger)^N\Bigr)\nn\\
                                \times&&\Bigl((\aop_{2L}^\dagger)^N+(\aop_{2R}^\dagger)^N\Bigr)
 \ket{0}.
\eea
To calculate the time evolution  $\ket{\Psi(t)}$ after the preparation time at $t=0$, 
we go beyond the perturbative interaction-picture method of ref. \cite{aziz2025classical} 
and work in the Heisenberg picture, switching to the Schrödinger picture at the end.

On a fixed background metric, the quantization of the Heisenberg field $\phiop(x)$ 
and of the total Hamiltonian $\hat{H}$ are standard \cite{fulling1989}. 
The field is expanded in the orthonormal basis of the mode functions $\varphi_\al$, 
which themselves are classical solutions of the covariant Klein--Gordon equation. 
If the background is static (as it will be in our case), the quantized
Heisenberg field admits the mode expansion (5), and the total Hamiltonian 
takes the diagonal form (6):
\bea
\phiop(x)&=&\sumint_\al 
\frac{1}{\sqrt{2\om_\al}}\aop_\al\varphi_\al(\xv)e^{-i\om_\al t}
\dots\\ 
\Hop&=&\sumint_\al \om_\al\aop_\al^\dagger\aop_\al+\dots~,
\eea
where $\aop_\al$ is the Fock annihilation operator of the mode $\varphi_\al$.
The ellipses stand for the anti-particle parts.

The Aziz and Howl model \cite{aziz2025classical} represents the background in the weak-field limit
of general relativity by the Newtonian potential $\Phi(\xv)$, 
which is determined by the initial state $\ket{\Psi}$, and considered
static for the whole duration of the calculated evolution $\ket{\Psi(t)}$. 
Accordingly, we consider the Heisenberg field $\phiop(x)$ for $t\geq0$, 
satisfying the Klein-Gordon equation in the static linearized gravity
$h_{\mu\nu}(x)=-2\Phi(\xv)\delta_{\mu\nu}$ used in ref. \cite{aziz2025classical}:
\beq
\partial_t^2\phiop-(1+4\Phi)\nabla^2\phiop+(1+2\Phi)m^2\phiop=0.
\eeq
The stationary solutions yield the mode functions 
$\varphi_\al(\xv)e^{-i\om_\al t}$ in the expansion (5) of the Heisenberg field.  
Eqs. (5)–(7) establish the general structure of the Heisenberg dynamics, 
but the arguments below do not rely on their detailed form; in particular, 
the definition (8) of the Fock operators is independent of the specific expansion (5).

Now we turn toward the construction and the dynamics of the state (4). 
We show that $\ket{\Psi(t)}$ remains unentangled. In the Heisenberg picture, 
our four Fock operators are defined as follows:
\beq
\aop_\ki=i\int \phit_\ki(\xv)\phiop^{(+)}(\xv,0)d^3\xv, 
\eeq
where $\phiop^{(+)}$ is the positive frequency part of $\phiop$.
These Fock operators define the state (4) in the Heisenberg picture. 
This Heisenberg state remains constant after its preparation at $t=0$, 
while the field $\phiop(x)$ evolves with the total Hamiltonian $\Hop$.
Now we switch to the Schrödinger picture, and evolve the state $\ket{\Psi}$ into
$\ket{\Psi(t)}=e^{-it\Hop}\ket{\Psi}$. 
Using eq. (4) and $\Hop\ket{0}=0$, the action of evolution operator $e^{-i\Hop t}$ is equivalent
to the unitary time-evolution $\aop_\ki(t)=e^{-i\Hop}\aop_\ki e^{-it\Hop}$ of the Fock operators.   
We get our central result in closed analytic form:
\bea
\ket{\Psi(t)}=\frac{1}{2N!}
&&\Bigl((\aop_{1L}^\dagger(t))^N+(\aop_{1R}^\dagger(t))^N\Bigr)\nonumber\\
\times&&\Bigl((\aop_{2L}^\dagger(t))^N+(\aop_{2R}^\dagger(t))^N\Bigr)
\ket{0}.
\eea
The effective factorized form is the following:
\bea
\ket{\Psi(t)}=\frac12&&\Bigl(\ket{N,t}_{1L}\ket{0}_{1R}+\ket{0}_{1L}\ket{N,t}_{1R}\Bigr)\nonumber\\
                    \otimes&&\Bigl(\ket{N,t}_{2L}\ket{0}_{2R}+\ket{0}_{2L}\ket{N,t}_{2R}\Bigr).
\eea
The time-dependent $N$-boson states are defined as follows:
\beq
\ket{N,t}_{\kappa i}=\frac{1}{\sqrt{N!}}(\hat{a}_{\kappa i}^\dagger(t))^N\ket{0}_{\kappa i}.
\eeq
The state $\ket{\Psi(t)}$ remains a product (unentangled) state (10),
contrary to the claim in ref. \cite{aziz2025classical}
but in accordance with what we know about the semiclassical theory of gravity,
where the classical gravity does not entangle the quantized matter.

Most recently, Gundhi at al. \cite{gundhi2026canclassical} revisited the 4$^{th}$ order 
perturbative derivation of Aziz and Howl, concluding that the reported entanglement
is an artifact of neglecting certain 4$^{th}$-order transition amplitudes.
Our exact non-perturbative derivation proves that the absence of such an entangling mechanism follows
directly from the structure of quantum field theory on a classical (even if $\ket{\Psi}$-dependent) background.
Contrary to the interesting concept of Aziz and Howl that classical gravity entangles second-quantized matter, 
the entangling mechanism simply does not exist, rendering its interpretation unnecessary.

\acknowledgements
I am gratefully  thank Chiara Marletto, Jonathan Oppenheim and Vlatko Vedral for useful discussions, 
Andrea Di Biagio for insightful critique.
This research was supported by the
National Research, Development and Innovation Office
''Frontline'' Research Excellence Program (Grant No.
KKP133827), and profited from the EU COST Actions (Grants CA23115, CA23130).


\end{document}